\documentclass[aps,prl,twocolumn,showpacs,floatfix,
   ,superscriptaddress]{revtex4}
\usepackage{graphicx} \usepackage{color} \usepackage{natbib}

\usepackage{hyperref}

\begin{document}

   \title{Neutrino signal of electron-capture supernovae
          from core collapse to cooling}

\author{L.~H\"udepohl} \affiliation{Max-Planck-Institut f\"ur
  Astrophysik, Karl-Schwarzschild-Str.~1, D-85748 Garching, Germany}
\author{B.~M\"uller} \affiliation{Max-Planck-Institut f\"ur
  Astrophysik, Karl-Schwarzschild-Str.~1, D-85748 Garching, Germany}
\author{H.-T.~Janka} \affiliation{Max-Planck-Institut f\"ur
  Astrophysik, Karl-Schwarzschild-Str.~1, D-85748 Garching, Germany}
\author{A.~Marek} \affiliation{Max-Planck-Institut f\"ur
  Astrophysik, Karl-Schwarzschild-Str.~1, D-85748 Garching, Germany}
\author{G.~G.~Raffelt} \affiliation{Max-Planck-Institut f\"ur
  Physik, Werner-Heisenberg-Institut, F\"ohringer Ring~6, D-80802
          M\"unchen, Germany}

\date{\today}

\begin{abstract}
An $8.8\,M_\odot$ electron-capture supernova was simulated in
spherical symmetry consistently from collapse through explosion to
essentially complete deleptonization of the forming neutron star.  The
evolution time ($\sim9\,$s) is short because high-density effects
suppress our neutrino opacities. After a short phase of
accretion-enhanced luminosities ($\sim$200$\,$ms), luminosity
equipartition among all species becomes almost perfect and the spectra
of $\bar\nu_e$ and $\bar\nu_{\mu,\tau}$ very similar, ruling out the
neutrino-driven wind as r-process site.  We also discuss consequences
for neutrino flavor oscillations.
\end{abstract}


\maketitle

{\em Introduction.}---During the first seconds after collapse, a
supernova (SN) core emits its binding energy, roughly 10\% of its
rest mass, in the form of neutrinos. In the delayed explosion
paradigm, supported at least for some progenitor stars by recent
simulations~\cite{Janka:2006fh}, neutrinos revive the stalled shock
wave and by their energy deposition explode the
star~\cite{Bethe:1984ux}. Later they drive a powerful wind and
through $\beta$-processes determine its role as a possible site for
r-process nucleo\-synthesis~\cite{Qian:1996xt}. Inevitable
deviations from spherical symmetry allow the neutrino flux to emit
gravitational waves~\cite{Epstein:1978dv} and to impart a
potentially large neutron-star recoil~\cite{Lai:2000pk}.

A sparse neutrino signal was observed from SN~1987A. Existing and
foreseen large detectors~\cite{Scholberg:2007nu} will operate for
decades, suggesting the next galactic SN will provide a
high-statistics signal and allow for a direct glance of its inner
workings. The cosmic diffuse neutrino background from all past SNe
(DSNB) is almost certainly detectable if gadolinium loading of
Super-Kamiokande succeeds~\cite{Watanabe:2008ru} or by future large
scintillator detectors~\cite{Autiero:2007zj}, pushing the frontiers
of neutrino astronomy to cosmic distances.

The fluxes and spectra differ for the species $\nu_e$, $\bar\nu_e$ and
$\nu_x$ (representing any of $\nu_{\mu,\tau}$ or
$\bar\nu_{\mu,\tau}$). Flavor oscillations swap
$\nu_e\leftrightarrow\nu_x$ and $\bar\nu_e\leftrightarrow\bar\nu_x$ in
part or completely, a process strongly affected by collective effects
and Mikheev-Smirnov-Wolfenstein resonances~\cite{Dighe:2008dq}. What
is seen in the neutrino-driven wind or by a detector thus depends not
only on what is emitted, but also on the matter profile and neutrino
mixing parameters.

Quantitative studies in these areas are impeded by large
uncertainties of the expected fluxes and spectra. This problem
partly derives from uncertainties of the explosion mechanism itself
and input physics such as the nuclear equation of state (EoS).
Significant variations are expected in dependence of the progenitor
mass, and sometimes rotation and magnetic fields may come into play.
However, even without such complications, the range of predictions
is large for the post-explosion cooling phase when by far most of
the neutrino loss happens.

The pioneering work of the Livermore group combined relativistic
hydrodynamics with multi-group three-flavor neutrino diffusion in
spherical symmetry (1D), simulating the entire evolution
self-consistently \cite{Totani:1997vj}. The spectra were hard over a
period of at least 10$\,$s with increasing \hbox{hierarchy}
$\langle\epsilon_{\nu_x}\rangle > \langle\epsilon_{\bar\nu_e}\rangle
> \langle\epsilon_{\nu_e}\rangle$. These models, however, included
significant numerical approximations and omitted neutrino reactions
that were later recognized to be important~\cite{Keil:2002in}. A
crucial ingredient to enhance the early neutrino fluxes was a
neutron-finger mixing instability, which today is
disfavored~\cite{bruenn96}.

Relativistic calculations of proto neutron star (PNS) cooling with a
flux-limited equilibrium~\cite{Burrows:1986me, Keil:1995hw} or
multi-group diffusion treatment~\cite{suzuki} found monotonically
decreasing neutrino energies after no more than a short
\hbox{($\alt100\,$ms)} period of increase. Pons et~al.\
\cite{Pons:1998mm} studied PNS cooling for different EoS and masses,
using flux-limited equilibrium transport with diffusion coefficients
adapted to the underlying EoS. They always found spectral hardening
over 2--5$\,$s before turning over to cooling.

New opportunities to study the neutrino signal consistently from
collapse to late-time cooling arise from the class of
``electron-capture SNe'' (ECSNe) or ``O-Ne-Mg core SNe.'' These low-mass
(8--$10\,M_\odot$) stars collapse because of rapid electron capture
on Ne and Mg and could represent up to 30\% of all
SNe~\cite{Wanajo:2008bw}. They are the only cases where 1D
simulations obtain neutrino-powered explosions~\cite{Kitaura:2005bt}
and 2D yields only minor dynamical and energetic
modifications~\cite{Janka:2007di}. It has become possible to
carry hydrodynamic simulations with modern neutrino Boltzmann
solvers in 1D all the way to PNS cooling.

Very recently, the Basel group has circulated first results of the
PNS evolution~\cite{Fischer:2009af} for a representative
$8.8\,M_\odot$ progenitor~\cite{nomoto} using Shen et~al.'s
EoS~\cite{Shen:1998gq}, which is relatively stiff and yields cold NS
radii around 15~km.

Here we present our own long-term simulations of the same progenitor
and the same EoS, facilitating a direct comparison (results
with different EoS will be reported elsewhere). We will show that
improved neutrino interaction rates lead to significant differences.


{\em Numerical method.}---Our simulations were performed with the
\textsc{Prometheus/Vertex} code. It couples an explicit third-order
Riemann-solver-based Newtonian hydrodynamics code with an implicit
multi-flavor, multi-energy-group two-moment closure scheme for
neutrino transport.  The variable Eddington-factor closure is
obtained from a model Boltzmann equation~\cite{Rampp:2002bq}. We
account for general relativistic (GR) corrections with an effective
gravitational potential (case~A of Ref.~\cite{Marek:2005if}) and the
transport includes GR redshift and time dilation. Tests showed good
overall agreement until several 100~ms after core
bounce~\cite{Liebendoerfer:2003es, Marek:2005if} with fully
relativistic simulations of the Basel group's
\textsc{Agile-Boltztran} code. A more recent comparison with a GR
program~\cite{mueller10} that combines the \textsc{CoCoNut} hydro
solver~\cite{Dimmelmeier:2002bk} with the \textsc{Vertex} neutrino
transport, reveals almost perfect agreement except for a few
quantities with deviations of at most \hbox{7--10\%} until several
seconds. The total neutrino loss of the PNS agrees with the
relativistic binding energy of the NS to roughly 1\%, defining the
accuracy of global energy and lepton-number conservation in our
simulations.

Our primary case (Model~Sf) includes the full set of neutrino
reactions described in Appendix~A of Ref.~\cite{Buras:2005rp} with
the original sources. In particular, we account for nucleon recoils
and thermal motions, nucleon-nucleon (NN) correlations, weak
magnetism, a reduced effective nucleon mass and quenching of the
axial-vector coupling at high densities, NN bremsstrahlung,
$\nu\nu$ scattering, and
$\nu_e\bar\nu_e\to\nu_{\mu,\tau}\bar\nu_{\mu,\tau}$. In addition,
we include electron capture and inelastic neutrino scattering on
nuclei~\cite{Langanke:2003ii}.

To compare with previous simulations and the Basel
work~\cite{Fischer:2009af} we also consider in Model~Sr a reduced 
set of opacities, omitting pure neutrino interactions and all mentioned
improvements of the neutrino-nucleon interactions relative to the 
treatment of~\cite{Bruenn:1985en}.  

\begin{figure}
\includegraphics[width=8.5cm]{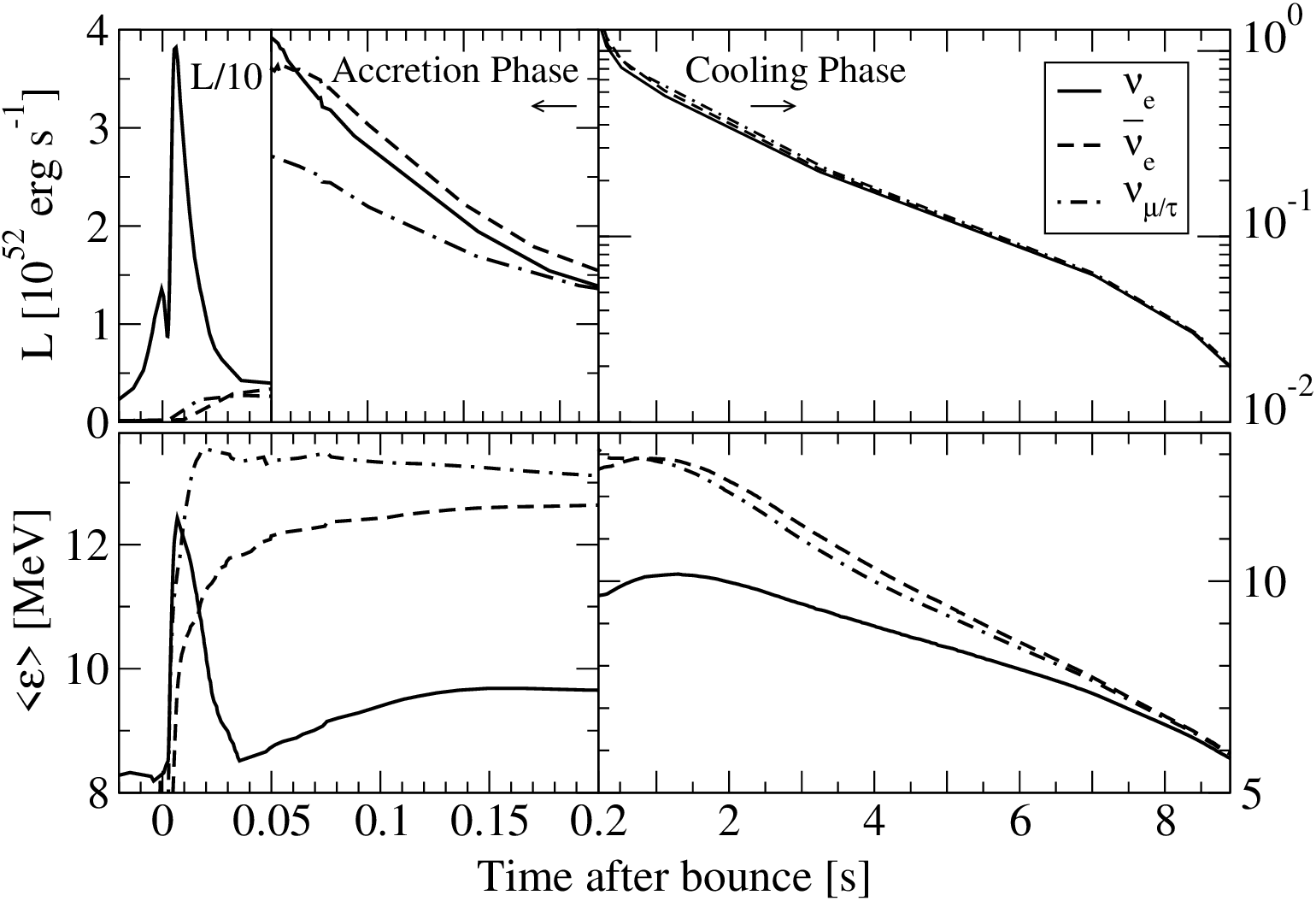}
\vskip12pt plus3pt minus3pt
\includegraphics[width=8.5cm]{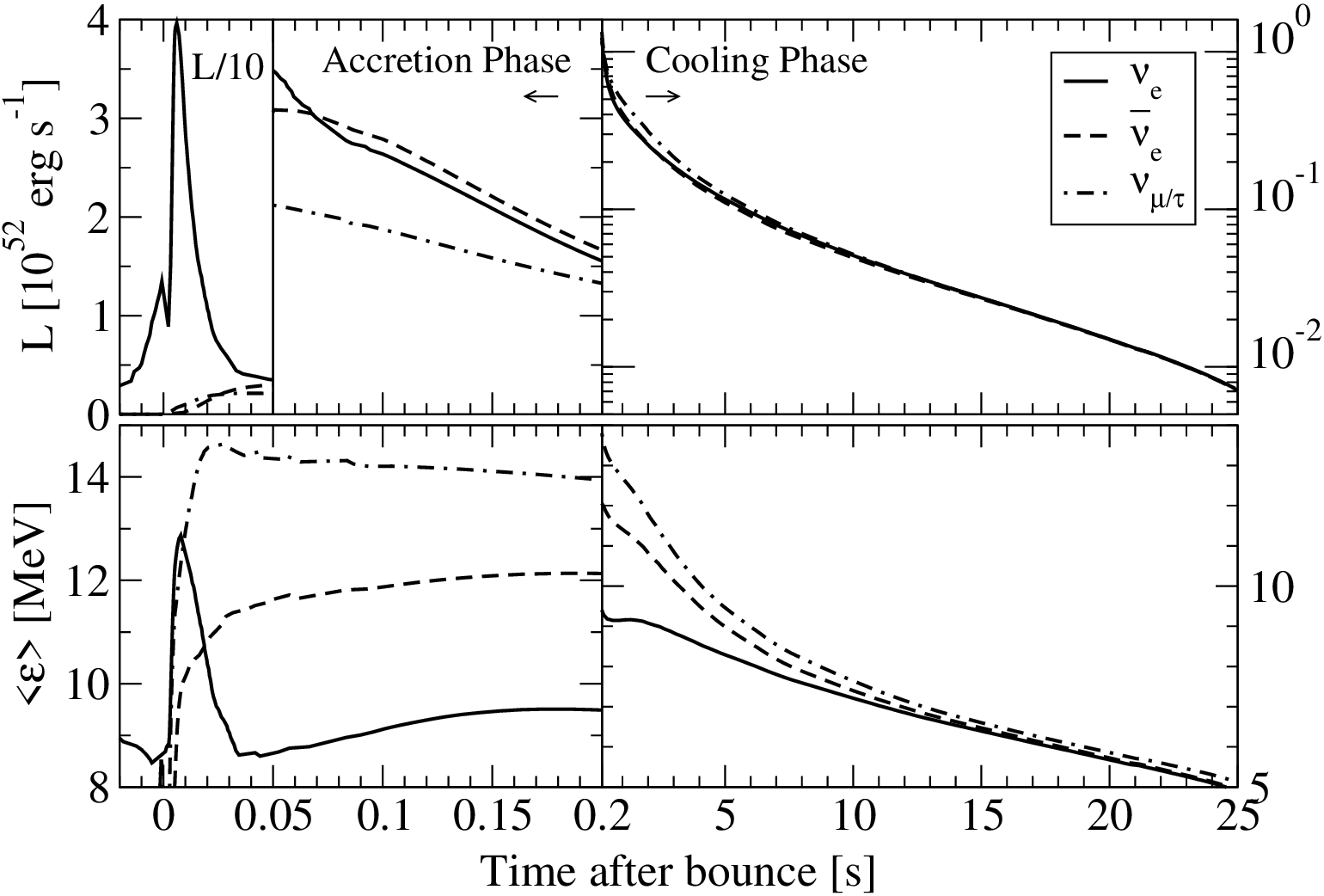}
\caption{
  Neutrino luminosities and mean energies observed at infinity.
  {\it Top:\/}~Full set of neutrino opacities (Model~Sf).
  {\it Bottom:\/}~Reduced set (Model~Sr).
  }
\label{fig:Lefull}
\end{figure}

{\em Long-term simulations.}---In Fig.~\ref{fig:Lefull} we show the
evolution of the $\nu_e$, $\bar\nu_e$ and $\nu_x$ luminosities and
of the average energies, defined as the ratio of energy to number
fluxes. The dynamical evolution, development of the explosion, and
shock propagation were previously described~\cite{Kitaura:2005bt,
Janka:2007di}. The characteristic phases of neutrino emission are
clearly visible: (i)~Luminosity rise during collapse. (ii)~Shock
breakout burst. (iii)~Accretion phase, ending already at $\sim$0.2$\,$s
post bounce when neutrino heating reverses the infall.
(iv)~Kelvin-Helmholtz cooling of the hot PNS with a duration of 10$\,$s
or more, accompanied by mass outflow in the neutrino-driven wind.

The PNS evolves in the familiar way~\cite{Burrows:1986me,
Pons:1998mm} through deleptonization and energy loss. It contracts,
initially heating up by compression and down-scattering of energetic
$\nu_e$ produced in captures of highly degenerate electrons. With
progressing neutronization the PNS cools, approaching a state of
$\beta$-equilibrium with vanishing $\nu_e$ chemical potential
$\mu_{\nu_e}$ and minimal electron content.

In Model~Sf, deleptonization and cooling take $\sim$10$\,$s until
$\nu$ transparency is approached. For $t>8.9\,$s we find $T \alt
6$~MeV and $\mu_{\nu_e}\sim 0$ throughout, and $\dot N_{\mathrm{L}}
\ll 10^{53}~{\rm s}^{-1}$. The final baryon mass is $M_\mathrm{b} =
1.366\,M_\odot$ with radius $\sim$15~km. Neutrinos have carried away
lepton number of $6.57\times 10^{56}$ and energy $E_\nu = 1.66\times
10^{53}\,$erg, so the gravitational mass is $M = M_\mathrm{b} -
E_\nu/c^2 = 1.273\,M_\odot$. The evolution is faster than in
previous works~\cite{Pons:1998mm} or in Model~Sr because the
high-density $\nu$
opacities are suppressed, where
NN correlations~\cite{Reddy:1997yr} probably dominate.
In Model~Sr, deleptonization continues at 25$\,$s
on the low level of  
$\dot N_{\mathrm{L}}\lesssim 10^{53}\,$s$^{-1}$, 
$T_{\rm center}\sim 11.5\,$MeV, and only 97\% of
the gravitational binding energy have been lost.

Differences are also conspicuous in the luminosities. Until 5.5$\,$s
they are higher (up to 60\% at $t\sim 2\,$s) in Model~Sf, whereas
afterwards they drop much faster compared to Model~Sr. On the other
hand, for $t\agt0.2\,$s, after the end of accretion, the
luminosities in both models become independent of flavor within 10\%
or better. The total radiated $E_\nu$ shows nearly equipartition:
20\% are carried away by $\nu_e$, 16\% by $\bar\nu_e$, and
4$\times$16\% by $\nu_x$.


{\em Spectra.}---The mean neutrino energies evolve
very differently in the two cases. While they increase over
\hbox{1--1.5$\,$s} for $\nu_e$ and $\bar\nu_e$ in Model~Sf, 
they increase
only until $\sim$0.2$\,$s in Model~Sr. The opacities are lower and
thus the neutrino spheres at higher $T$, so Model~Sf has larger
$\langle\epsilon_{\nu_e}\rangle$ and
$\langle\epsilon_{\bar\nu_e}\rangle$ for several seconds before
dropping below Model~Sr due to the faster overall evolution.

The canonical spectral hierarchy $\langle\epsilon_{\nu_x}\rangle >
\langle\epsilon_{\bar\nu_e}\rangle > \langle\epsilon_{\nu_e}\rangle$
persists in Model~Sr during the cooling phase, while in Model~Sf we 
find $\langle\epsilon_{\nu_x}\rangle \approx
\langle\epsilon_{\bar\nu_e}\rangle > \langle\epsilon_{\nu_e}\rangle$.
The close similarity and actually slight cross over of 
$\langle\epsilon_{\nu_x}\rangle$ 
and $\langle\epsilon_{\bar\nu_e}\rangle$ is caused by $\nu$ energy
transfer in $\nu N\to N \nu$. As recognized 
previously~\cite{Keil:2002in}, this in particular suppresses the
high-energy tail of the $\nu_x$ spectrum (spectral pinching) and
reduces $\langle\epsilon_{\nu_x}\rangle$ because the $\nu_x$ 
energy sphere is at higher density and surrounded by a thick 
scattering layer.

A quasi-thermal spectrum can be characterized by its lowest energy
moments \hbox{$\bar{\epsilon} \equiv \langle\epsilon_{\nu}\rangle$}
and $\langle\epsilon_{\nu}^2\rangle$. Simple analytic fits use a
nominal Fermi-Dirac (FD) distribution with temperature $T$ and
degeneracy parameter $\eta$~\cite{janka89} or a modified power law
$f_\alpha(\epsilon) = (\epsilon/\bar{\epsilon})^\alpha {\rm
e}^{-(\alpha+1)\epsilon/\bar{\epsilon}}$ ~\cite{Keil:2002in}.  The
spectrum is ``pinched'' (narrower than a thermal FD) if $p =
a^{-1}\langle\epsilon_{\nu}^2\rangle/
\langle\epsilon_{\nu}\rangle^2=a^{-1}(2+\alpha)/(1+\alpha) <1$ where
$a \approx 1.3029$. So it is pinched for $p<1$, $\eta>0$ and
$\alpha\agt2.3$ and anti-pinched otherwise. A Maxwell-Boltzmann (MB)
spectrum has $\alpha=2$, $\eta = -\infty$ and $p\approx 1.0234$.

\begin{figure}
\includegraphics[width=8.5cm]{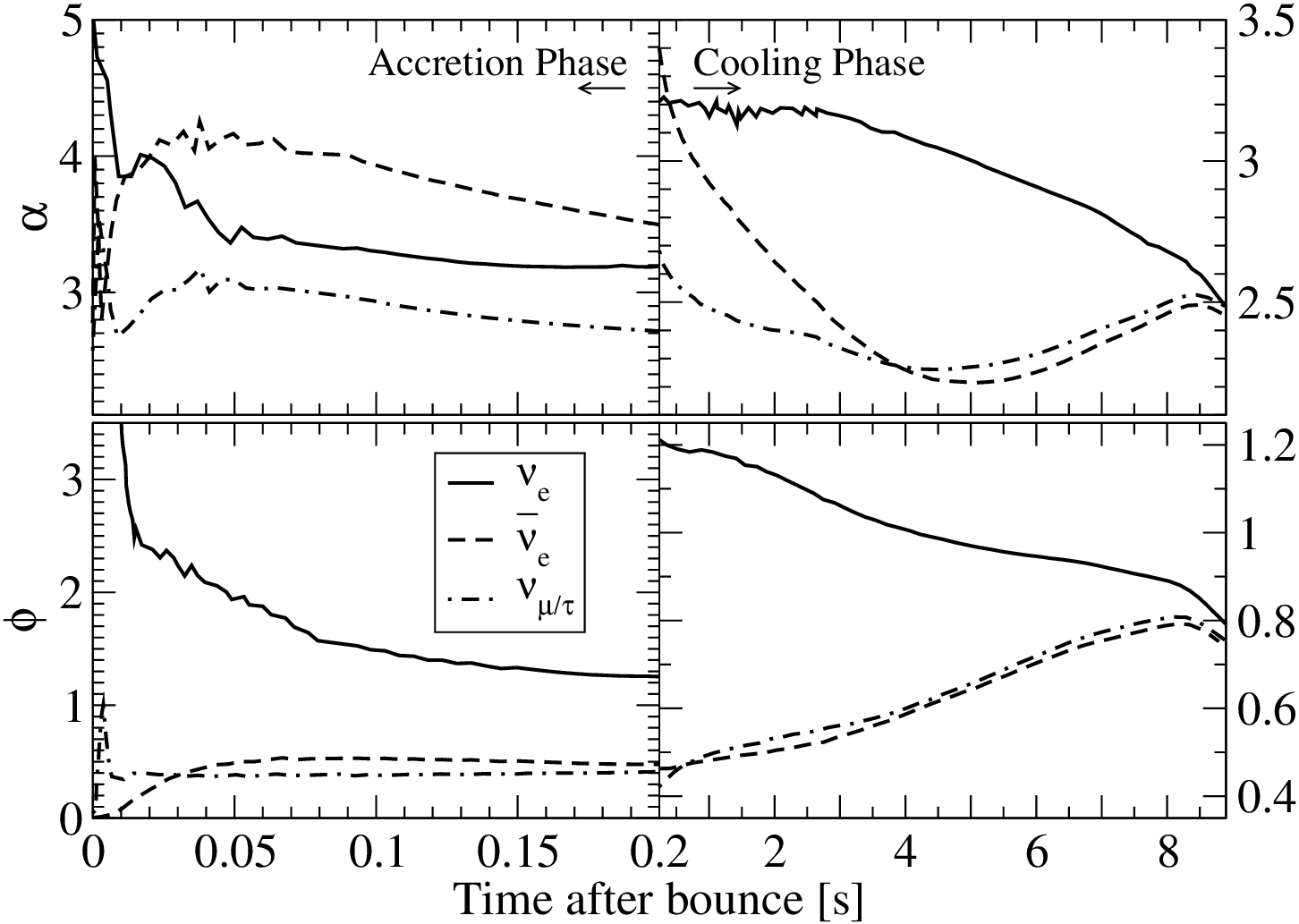}
\caption{Spectral fit parameter $\alpha$ and luminosity grayness
  factor $\phi$ for Model~Sf.}
\label{fig:spectrapar}
\end{figure}

In Model~Sf the $\nu_e$ spectrum is always pinched, while $\bar\nu_e$
and $\nu_x$ are mildly anti-pinched ($-1 < \eta < 0$) for $3\,{\rm s}
\alt t\alt 7\,$s (Fig.~\ref{fig:spectrapar}). At the end of the
simulation $\langle\epsilon\rangle$ becomes almost identical for all
species. The same applies to the spectral shape, which approaches a
thermal FD function ($\alpha \approx 2.5$, $p \approx 0.99$, $\eta
\approx 0.6$).

The time-integrated spectra of the number fluxes have
$\langle\epsilon_{\nu_e,\bar\nu_e,\nu_x}\rangle = 9.40$, 11.44, and
11.44~MeV. The spectrum is moderately pinched for $\nu_e$ ($p \approx
0.96$, $\alpha \approx 3.0$, $\eta\approx 1.7$), a nearly thermal FD
for $\bar\nu_e$ ($p \approx 0.99$), and slightly anti-pinched for
$\nu_x$ ($p \approx 1.017$, $\alpha \approx 2.1$, $\eta \approx
-1.5$).


{\em Effective radiating surface.}---The neutrino luminosities
$L_\nu$ and effective temperatures $T_{\rm e}$ can be used to
estimate the NS circumferential radius $R$. (This does
not apply to late-time volume emission and the early accretion-powered phase.)
The Stefan-Boltzmann law is $L_\nu
= 4\pi\phi \sigma_\nu T_\mathrm{e}^4 R_\infty^2$ in terms of
quantities measured at infinity and $\sigma_\nu=4.751\times
10^{35}\,{\rm erg}\,{\rm MeV}^{-4}\,{\rm cm}^{-2}\,{\rm s}^{-1}$ if
$T_\mathrm{e}$
is measured in MeV. A MB spectrum is assumed ($T_\mathrm{e}
=\frac{1}{3}\, \langle\epsilon_{\nu}\rangle$) with isotropic
emission at the radiating surface. All deviations from these
assumptions are absorbed in a ``grayness factor'' $\phi$.

We define $R$ as the location where $\rho=10^{11}~{\rm
  g}\,{\rm cm}^{-3}$. GR corrections imply that $R_\infty = R/
\sqrt{1-2\beta}$ where $\beta=GM/(Rc^2)$ and $M$ is the PNS
gravitational mass. In Model~Sf, $R/R_\infty$ drops from an initial
value near 1 to 0.88 at 3$\,$s, followed by a slow decline to 0.87
at 8$\,$s. $M$ is linked to the gravitational binding energy and
thus to the total $\nu$ energy release by $E_\nu \approx
0.6\beta Mc^2(1-0.5\beta)^{-1}$~\cite{Lattimer:2000nx}, reproduced
very well in our simulations.  We propose to use these relations
with measured values of $E_\nu$,
$\langle\epsilon_{\bar\nu_e}\rangle$ and $L_{\bar\nu_e}$ during the
cooling phase to determine $M$ and $R$ from the signal of a future
galactic~SN.

The grayness factors for Model~Sf are shown in
Fig.~\ref{fig:spectrapar}. Their time variation is considerable, but
$\phi\approx 0.6$ is a good choice for $\bar\nu_e$
around the time (5--6$\,$s) of the $\alpha(t)$ minimum. This estimate
applies for different EoS that we have tested. 
We also found that the evolutions of $\alpha(t)$ and $\phi(t)$
are not sensitive to the EoS.

\begin{figure}[b]
\includegraphics[width=8.5cm]{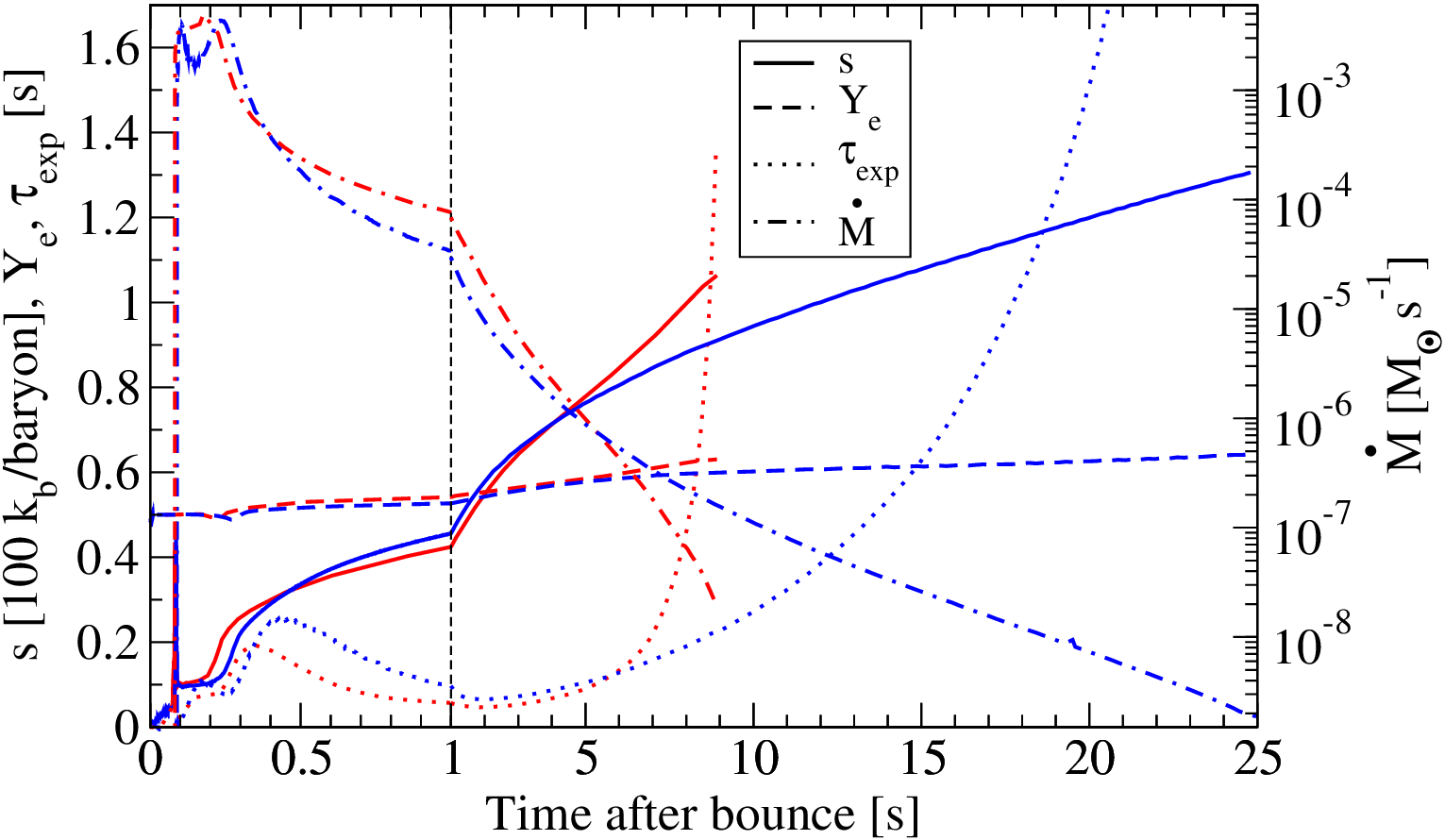}
\caption{ Neutrino-driven wind properties (Model~Sf: red, 
  Model~Sr: blue). 
  $\tau_{\rm exp}$ refers to the $T$-decline at 0.5$\,$MeV.
  NOTE THE ERRATUM ON PAGE \pageref{page:erratum}.
  \label{fig:wind}}
\end{figure}

{\it Neutrino-driven wind}.---Absorption of $\nu_e$ and $\bar\nu_e$ on
nucleons determines the $n/p$ ratio $Y_e^{-1} - 1$ in the
neutrino-driven wind~\cite{Qian:1996xt}, where $Y_e$ is the
electron/baryon ratio. R-process nucleosynthesis conditions depend on
$n/p$ in addition to the entropy per baryon $s$, the expansion
timescale $\tau_\mathrm{exp}$ (between $T = 0.5\,$MeV and
$0.5/e\,$MeV), and the mass-loss rate $\dot M$, which
in turn depend on the neutrino energy
deposition~\cite{Qian:1996xt}.

Since $\langle\epsilon_{\nu_e} \rangle$ and
$\langle\epsilon_{\bar\nu_e}\rangle$ are very similar and because
$\bar\nu_e$ absorptions are impeded by the $n/p$ mass difference,
one finds $Y_e$ values significantly above 0.5~\cite{Fischer:2009af}.
We confirm this result in our Model~Sf (Fig.~\ref{fig:wind}). The 
mean energies approach each other at late times, so $Y_e$ grows
monotonically and reaches $\sim$0.63 after 9$\,$s.
Proton excess was suspected 
earlier~\cite{Horowitz:1999} and in addition
to insufficient entropies~\cite{Thompson:2001} disfavors 
r-processing even in the late wind.
Flavor conversions of active neutrinos cannot change this 
conclusion because $\langle\epsilon_{\bar\nu_e}\rangle\sim
\langle\epsilon_{\bar\nu_x}\rangle$.


{\it Conclusions.}---Our simulations of an ECSN
confirm that the difference between
$\langle\epsilon_{\bar\nu_e}\rangle$ and $\langle\epsilon_{\nu_e}\rangle$
is too small for n-excess in the $\nu$-driven wind, excluding ECSNe
as r-process site~\cite{Fischer:2009af}. When nucleon recoils are
included (Model~Sf), the mild hierarchy 
$\langle\epsilon_{\nu_x}\rangle > \langle\epsilon_{\bar\nu_e}\rangle
> \langle\epsilon_{\nu_e}\rangle$ found in~\cite{Fischer:2009af} and
in our Model~Sr changes to $\langle\epsilon_{\nu_x}\rangle
\approx \langle\epsilon_{\bar\nu_e}\rangle 
> \langle\epsilon_{\nu_e}\rangle$. Thus flavor
conversions in the $\bar\nu$ sector would hardly have any impact.

The PNS cooling time is significantly shortened by
high-density nuclear effects in the neutrino opacities.
A steep density but shallow $T$
profile near the PNS surface causes $\nu_e$, $\bar\nu_e$, and 
$\nu_x$ to be radiated from a thermal bath with similar
neutrinospheric radii and temperatures. Luminosity equipartition
among all species during the cooling phase is therefore almost
perfect, compatible with~\cite{Totani:1997vj} and \cite{Fischer:2009af} and 
despite different spectral hierarchies. Only 
during accretion-powered neutrino emission is $L_{\nu_e,\bar\nu_e}$ 
significantly larger than $L_{\nu_x}$, and flavor oscillations
would most easily show up in a high-statistics SN $\bar\nu_e$
signal during this phase. Differences between the $\nu_e$ and $\nu_x$
fluxes and spectra are pronounced in all phases. Therefore, a large
$\nu_e$ detector would be especially useful~\cite{Lunardini:2008}.

The time-integrated 
$\langle\epsilon_{\bar\nu_e,\bar\nu_x}\rangle=11.4$~MeV
is relatively low.
Results in~\cite{Fischer:2009af} suggest that
$\langle\epsilon_{\bar\nu_e,\bar\nu_x}\rangle=11$--12~MeV may be
typical 
also for more massive progenitors and PNS.
If so, the agreement with the SN1987A $\nu$ data would be much better
than previously thought~\cite{Jegerlehner:1996kx}.

Our ECSN simulations with different softer and
stiffer EoS (to be published elsewhere) yield similar
results and corroborate our conclusions. The time-integrated
$\langle\epsilon_{\bar\nu_e,\bar\nu_x}\rangle$
differs by no more than 
$\sim$0.5~MeV.

Emission differences of $\nu_e$ and $\bar\nu_e$ and 
wind properties 
depend only modestly on the PNS mass up to nearly the black hole
limit~\cite{Pons:1998mm,Qian:1996xt,Thompson:2001}. 
PNS winds with p-excess thus probably disfavor r-processing also in
other SNe, as already seen for 10.8 and 18$\,M_\odot$ stars
in~\cite{Fischer:2009af}. Our simulations with PNS convection 
(to be reported elsewhere) also yield $Y_e > 0.5$, whereas
$Y_e\lesssim 0.3$ is needed for a strong r-process with typical
$s$ and $\tau_\mathrm{exp}$ values obtained in wind 
models~\cite{Hoffman:1996aj,Thompson:2001}.
It remains to be explored if $Y_e$ in common SNe
can be sufficiently reduced by a new physical
mechanism, perhaps involving rotation, magnetic fields or a modified
composition (e.g.\ light clusters~\cite{Arcones:2008kv}).

\begin{acknowledgments}
We acknowledge partial support by DFG grants No.\ SFB/TR~7,
SFB/TR~27, EXC~153 and computer time at the HLRS in Stuttgart
and NIC in J\"ulich.
\end{acknowledgments}



\clearpage

\section*{Erratum}

\label{page:erratum}

In the original paper, a factor of $4 \pi$ was incorrectly omitted
in computing the mass-loss rates $\dot{M}$ plotted in Fig.~3. The
corrected version of Fig.~3 is given here. This correction does not
affect the conclusions of the Letter. No reference was actually made
to numerical values of $\dot{M}$ in the text.

\setcounter{figure}{2}
\begin{figure}[h]
\includegraphics[width=8.5cm]{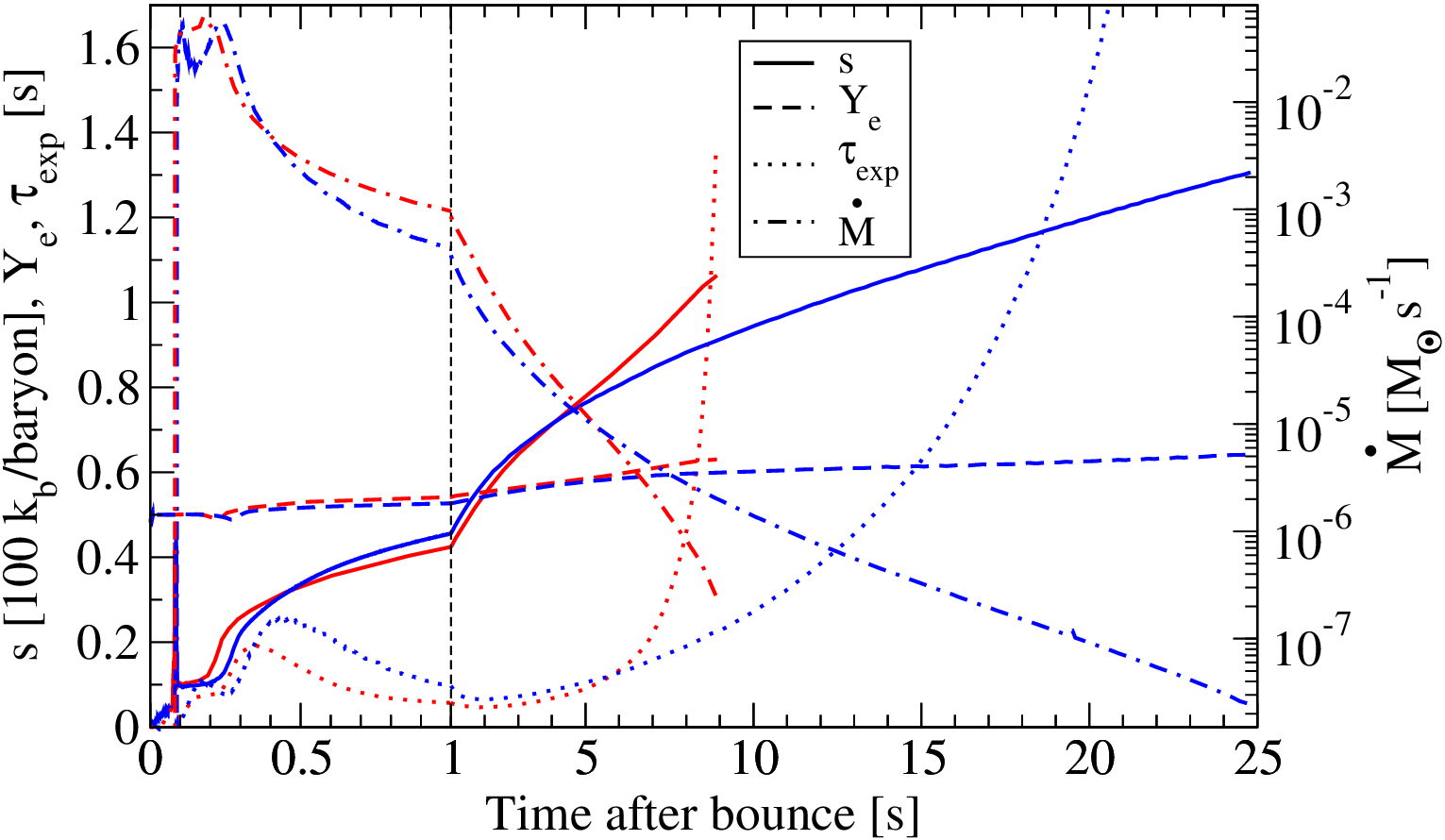}
\caption{Neutrino-driven wind properties (Model~Sf: red, 
  Model~Sr: blue). 
  $\tau_{\rm exp}$ refers to the $T$-decline at 0.5$\,$MeV.
  \label{fig:wind_new}}
\end{figure}

\end{document}